\documentstyle[emulateapj,times,psfig]{article}

\def\gta{\ifmmode {\mathbin{\lower 3pt\hbox   
    {$\,\rlap{\raise 5pt\hbox{$\char'076$}}\mathchar"7218\,$}}}
    \else {${\mathbin{\lower 3pt\hbox
    {$\rlap{\raise 5pt\hbox{$\char'076$}}\mathchar"7218\,$}}}
    $}\fi}
\def\lta{\ifmmode {\,\mathbin{\lower 3pt\hbox   
    {$\,\rlap{\raise 5pt\hbox{$\char'074$}}\mathchar"7218\,$}}}
    \else {${\mathbin{\lower 3pt\hbox
    {$\rlap{\raise 5pt\hbox{$\char'074$}}\mathchar"7218\,$}}}
    $}\fi}

\begin{document}

\title{Four-Body Effects in Globular Cluster Black Hole Coalescence}

\author{M. Coleman Miller and Douglas P. Hamilton}
\affil{Department of Astronomy, University of Maryland\\
       College Park, MD  20742-2421\\
       miller@astro.umd.edu,hamilton@astro.umd.edu}

\begin{abstract}
In the high density cores of globular clusters, multibody interactions
are expected to be common, with the result that black holes in
binaries are hardened by interactions. It was shown by Sigurdsson \&
Hernquist (1993) and others that $10\,M_\odot$ black holes interacting
exclusively by three-body encounters do not merge in the clusters
themselves, because recoil kicks the binaries out of the clusters
before the binaries are tight enough to merge.  Here we consider a new
mechanism, involving four-body encounters.  Numerical simulations by a
number of authors suggest that roughly 20-50\% of binary-binary
encounters will eject one star but leave behind a stable hierarchical
triple.  If the orbital plane of the inner binary is strongly tilted
with respect to the orbital plane of the outer object, a secular Kozai
resonance, first investigated in the context of asteroids in the Solar
System, can increase the eccentricity of the inner body significantly.
We show that in a substantial fraction of cases the eccentricity is
driven to a high enough value that the inner binary will merge by
gravitational radiation, without a strong accompanying kick.
Thus the merged object remains in the cluster; depending on the binary
fraction of black holes and the inclination distribution of
newly-formed hierarchical triples, this mechanism may allow massive
black holes to accumulate through successive mergers in the cores of
globular clusters.  It may also increase the likelihood that stellar-mass
black holes in globular clusters will be detectable by their gravitational
radiation.

\end{abstract}


\section{Introduction}

Globular clusters are outstanding testbeds for dynamics.  As dense systems
with ages many times their core relaxation time, they display such features as
core collapse and mass segregation, and they are almost certainly affected
strongly by the presence of even a small number of binaries.  It has long
been speculated that various processes might produce relatively massive
black holes in their cores (e.g., Wyller 1970; Bahcall \& Ostriker 1975;
Frank \& Rees 1976; Lightman \& Shapiro 1977; Marchant \& Shapiro 1980;
Quinlan \& Shapiro 1987; Portegies Zwart et al. 1999; Ebisuzaki et al.
2001). Recent observations of some dense clusters provide tentative
evidence for black holes as massive as $2500\,M_\odot$ at their centers
(Gebhardt et al. 2000).

Qualitatively, it seems entirely reasonable that large black holes
should grow in the cores of many clusters.  Even at birth, black holes
are much more massive than the average star in a cluster, and hence
they sink rapidly towards the core.  When in the core, they tend to
exchange into binaries.  If the binary is hard (i.e., if its binding
energy exceeds the average kinetic energy of a field star), then a
subsequent interaction with a field star tends to harden the binary
(e.g., Heggie 1975).
If this process is repeated often enough, the binary becomes tight enough
that it can merge by gravitational radiation, and the black hole becomes
larger.  If these binaries merge while still in the cluster, sources in
globulars could be excellent prospects for detection by the upcoming
generation of gravitational wave instruments.

However, it has been shown (e.g., Sigurdsson \& Hernquist 1993;
Portegies Zwart \& McMillan 2000) that if all black holes have initial
masses of $10\,M_\odot$, three-body encounters alone do {\it not} lead to the
formation of a large black hole at the center.  The reason is that
hardening in a binary-single interaction is accompanied by recoil,
which kicks the binary out of the cluster before it can merge.  Without
additional effects, this means that the mergers occur well away from
their host globulars.  If the
initial mass of a black hole is $\gta 50\,M_\odot$, as may result from
a high-mass low-metallicity star or rapid merger of main sequence
stars, it has enough inertia to remain in the core and grow by
coalescence (Miller \& Hamilton 2002).  But what if only low-mass
black holes are produced?

Here we propose a new mechanism for the coalescence of low-mass black holes in
globular clusters, involving binary-binary interactions.  Studies of such
four-body encounters have been comparatively rare, but have shown that in
roughly 20-50\% of the interactions the final state is an unbound single
star plus a stable hierarchical triple system (Mikkola 1984; McMillan,
Hut, \& Makino 1991; Rasio, McMillan, \& Hut 1995).  This allows an
important new effect: studies of planetary and stellar systems have shown
that if there is a large relative inclination between the orbital planes
of the inner binary and the outer object of the triple, then over many
orbital periods the relative inclination periodically trades off with the
eccentricity of  the inner binary, sometimes leading to very high
eccentricities (Kozai 1962; Harrington 1968, 1974; Lidov \& Ziglin 1976).  In
turn, this can enhance the gravitational radiation rate enormously,
leading to merger without a strong kick and allowing even low-mass binary
black holes in globulars to be potential gravitational wave sources.

In \S~2 we discuss the principles of this resonance, as derived in the
case of three objects of arbitrary mass by Lidov \& Ziglin (1976).  To
their treatment we add, in \S~3, a simple term that accounts for
general relativistic pericenter precession.  We show that although, as
expected, this precession decreases the maximum attainable
eccentricity for a given set of initial conditions, the decrease is
typically minor and thus there is
significant phase space in which the eccentricity resonance leads to
rapid merger.  In \S~4 we use these results in a simple model for the
mergers of black holes, and show that, depending on the fraction of
black holes in binaries, this effect can lead to a dramatic increase
in the retention of black holes in globulars, and to the growth of
$\sim 10^{2-3}\,M_\odot$ black holes in their cores.

\section{Principles of the Kozai Resonance}

When looking for changes in the orbital properties of a three-body
system that extend over many orbital periods of both the inner binary
and the outer tertiary, it is convenient to average the motion over
both these periods, a procedure called double averaging.  A general
analysis of the double-averaged three-body problem has been performed
to quadrupolar order for Newtonian gravity by Lidov \& Ziglin (1976)
in Hill's case, in which the distance of the outer object (of mass
$m_2$) from the inner binary (with component masses $m_0$ and $m_1\leq
m_0$) is much greater than the semimajor axis of the inner binary.
They find that for any set of three masses there is always a relative
inclination of orbits such that an inner binary with arbitrarily small
initial eccentricity will evolve to $e=1$.  For example, in the
restricted three-body problem in which $m_0\gg m_2\gg m_1$ (e.g., the
Sun, Jupiter, and an asteroid interior to Jupiter's orbit; see Kozai
1962), a relative orbital inclination of 90$^\circ$ will cause the
asteroid to evolve to $e=1$ in a finite time.

However, the growth to such high eccentricity depends on a long series
of perturbations from the tertiary that add coherently, and hence
requires certain phase relations.  An extra source of precession of
the pericenter can interfere with this.  For example, the orbits of
the moons of Uranus are tipped by 97$^\circ$ with respect to Uranus'
orbit around the Sun, but their eccentricities stay relatively low due
to precession introduced by the quadrupole moment of Uranus.  In the
case of black holes or other close massive objects, a similar role may
be played by the effects of general relativity, which to lowest order
includes precession of the pericenter.  How does this affect the
maximum eccentricity for a given set of initial conditions?

Hill's approximation allows us to treat the system as two nested
binaries: the inner pair composed of $m_0$ and $m_1$, and a second
pair consisting of i) an object of mass $m_0+m_1$ located at their
center of mass and ii) $m_2$. Defining variables as in Lidov \& Ziglin
(1976), we let $M_1=m_0+m_1$ and $M_2=m_0+m_1+m_2$ be the total masses
of the two binaries, and $\mu_1=m_0m_1/M_1$, and $\mu_2=m_2M_1/M_2$ be
their reduced masses. Let the semimajor axes and eccentricities of the
two binaries be $a_1$, $e_1$ and $a_2$, $e_2$, and define $i_1$ and
$i_2$ to be the inclinations of the binaries relative to the invariant
plane of angular momentum of the system.  Finally, let
$\mu=Gm_0m_1m_2/M_1$ and $\epsilon=1-e_1^2$.

The double-averaged Hamiltonian ${\bar{\cal H}}$ admits several
integrals each of which yields a constant of the motion.  First, the
double-averaging procedure guarantees that $a_1$ and $a_2$ are
constant. We keep terms in the Hamiltonian up to linear order in
$a_1/a_2$; these quadrupolar terms dominate the evolution of the
system for the high relative inclinations of interest here (see Ford,
Kozinsky, \& Rasio 2000). To this order, $e_2$ is also constant. The
problem has two constants of the motion that are related to
angular momentum: 

\begin{equation}
\alpha=\epsilon^{1/2}\cos i_1+\beta\cos i_2\; ,\qquad
\beta={\mu_2\sqrt{M_2}\over{\mu_1\sqrt{M_1}}}\sqrt{{a_2\over a_1}
\left(1-e_2^2\right)}\; .
\end{equation}

The constant $\beta$ (a combination of the constants $a_1,a_2$, and
$e_2$) represents the total angular momentum of the outer binary while
$\alpha$ is the total system angular momentum (with contributions from
both the inner and outer binaries). Both $\alpha$ and $\beta$ are made
dimensionless by dividing by $L_1=\mu_1\sqrt{GM_1a_1}$, the angular
momentum that the inner binary would have if it were on a circular
orbit.

The Hamiltonian, ${\bar{\cal H}}$, itself is constant. For
convenience, we define ${\bar{\cal H}}=-k(W+{5\over 3})$, with $k=3\mu
a_1^2/\left[8a_2^3(1-e_2^2)^{3/2}\right]$ and obtain:
\begin{equation}
\label{eqn:W}
W=-2\epsilon+\epsilon\cos^2I+
5(1-\epsilon)\sin^2\omega\left(\cos^2I-1\right)\; ,
\end{equation}
which is equation~(30) from Lidov \& Ziglin (1976).  Here $\omega$ is the
argument of pericenter of the inner binary and the scaled angular
momenta $\alpha$, $\beta$, and $\sqrt{\epsilon}$ form a triangle from which
the relative inclination $I=i_1+i_2$ can be obtained using the
law of cosines:
\begin{equation}
\label{eqn:cosI}
\cos I={\alpha^2-\beta^2-\epsilon\over{2\beta\sqrt{\epsilon}}}\; .
\end{equation}
The maximum $\epsilon$ (and hence minimum $e_1$) occurs for
$\omega=0$, and the minimum $\epsilon$ (and hence maximum $e_1$)
occurs for $\omega=\pi/2$; see Lidov \& Ziglin (1976).  Therefore,
given initial values for $\epsilon_0$ and $\omega_0$, the maximum
eccentricity may be derived from conservation of $W$ at
$\omega=\pi/2$. The time required to push the system from its minimum
to maximum eccentricity, is of order
\begin{equation}
\tau_{\rm evol}\approx f\left({M_1\over m_2}{b_2^3\over a_1^3}\right)^{1/2}
\left(b_2^3\over{Gm_2}\right)^{1/2}
\end{equation}
(e.g., Innanen et al. 1997), where $b_2=a_2\sqrt{1-e_2^2}$ is the
semiminor axis of the tertiary and typically $f\sim{\rm few}$ for
$I$ near $90^\circ$, which is the case of interest here.

\section{The Kozai Resonance with GR Precession}

Post-Newtonian precession may be included in a couple of equivalent
ways.  One is to modify the Hamiltonian directly, by changing the
gravitational potential to simulate some of the effects of general
relativity.  The modification of the potential is not unique,
and depends on which aspect of general relativity is to be reproduced
(see Artemova, Bjornsson, \& Novikov 1996).  For our purposes it is the
precession of pericenter that is important (as opposed to, e.g., the
location of the innermost stable circular orbit), and hence the correct
lowest-order modification is $-GM/r\rightarrow (-GM/r)(1+3GM/rc^2)$
(Artemova et al. 1996).  Averaging the correction term over the orbits
of the tertiary and inner binary, we obtain a correction to the 
double-averaged Hamiltonian of
\begin{equation}
\label{eqn:HPN}
{\bar{\cal H}}_{\rm PN}
=-{3(Gm_0)^2m_1\over{a_1^2c^2\epsilon^{1/2}}}=-kW_{\rm PN}\; .
\end{equation}
This result may also be obtained from the first-order general
relativistic precession rate of $d\omega=(6\pi
GM_1/\left[a_1(1-e_1^2)c^2\right]$ over one binary period (see Misner,
Thorne, \& Wheeler 1973, p. 1110) using the equation of motion
$d\omega/dt=-(2k\sqrt{\epsilon}/L_1)(\partial W/\partial\epsilon)$
derived by Lidov \& Ziglin (1976). Substituting and integrating, we
find that the first-order post-Newtonian contribution to $W$ is
\begin{equation}
W_{\rm PN}={8\over\sqrt{\epsilon}}
{M_1\over m_2}\left(b_2\over{a_1}\right)^3{GM_1\over{a_1 c^2}}
\equiv \theta_{\rm PN}\epsilon^{-1/2}\; .
\end{equation}
in agreement with equation~(\ref{eqn:HPN}). We have also checked our
expressions with direct numerical three-body integrations; note that
equation~(\ref{eqn:HPN}) corrects a factor of two error in equation~(19) of
Lin et al. (2000).

Adding the new term $W_{\rm PN}$ to equation~(\ref{eqn:W}) and making use of
equation~(\ref{eqn:cosI}), we find
\begin{equation}
\label{eqn:Wgr}
W=-2\epsilon+\epsilon\cos^2I+5(1-\epsilon)\sin^2\omega(\cos^2I-1)
+\theta_{\rm PN}/\epsilon^{1/2} \; .
\label{eqn:ecc}
\end{equation}
As in the previous section, for a given set of initial conditions, one
can therefore solve for the minimum $\epsilon$ (maximum $e$), by
setting $\omega=\pi/2$ and using the conservation of $W$.
In general we expect that initially
the inner binary will have significant eccentricity caused by
perturbations during the four-body encounter, but for simplicity
we will assume that the initial eccentricity is small enough that
$\epsilon_0\approx 1$.  In the restricted three-body problem in which
$m_0\gg m_2\gg m_1$ and the initial relative inclination is $I_0$, the
approximate solution for $\epsilon_{\rm min}$ when $5\cos^2I_0\ll 3$
(high inclination) and $\theta_{\rm PN}\ll 3$ (weak precession) is
\begin{equation}
\label{eqn:emin}
\epsilon_{\rm min}^{1/2}\approx {1\over 6}\left[\theta_{\rm PN}+
\sqrt{\theta_{\rm PN}^2+60\cos^2I_0}\right]\; .
\end{equation}
When $60\cos^2I_0\gg\theta_{\rm PN}$ this reduces to the Newtonian
solution, in which the maximum eccentricity is $e_{\rm max}=
\sqrt{1-(5/3)\cos^2I_0}$ (Innanen et al. 1997).  If instead $I_0\approx
\pi/2$ so that $60\cos^2I_0\ll \theta_{\rm PN}$, then $e_{\rm
max}\approx 1-\theta_{\rm PN}^2/9$.  More generally, 
for any set of masses, if $e\rightarrow 1$ is allowed in the Newtonian
problem then $e_{\rm max}=1-{\cal O}(\theta_{\rm
PN}^2)$ when general relativistic precession is included.  
Numerically, for $M_1=M_\odot$ and $a_1=$1~AU,
$\theta_{\rm PN}=8\times 10^{-8}(M_1/m_2)(b_2/a_1)^3$.
Equation~(\ref{eqn:emin}) shows that in the restricted three-body
problem the maximum possible
eccentricity (minimum $\epsilon_{\rm min}$) is attained for the initial
condition $I_0 = \pi/2$ (initially perpendicular circular orbits). If
$m_1$ has non-negligible mass, so that $m_2$ dominates the total
angular momentum less, then the critical $I_0$ increases (Lidov \&
Ziglin 1976).  Figure~1 shows the critical inclination in the Newtonian
case ($\theta_{\rm PN}=0$) for several mass ratios and semimajor axes.

We want to know whether this process can cause the inner binary to
reach a high enough eccentricity that it merges by gravitational
radiation before the next encounter with a star in the globular
cluster (which will typically alter the eccentricities and
inclinations significantly).  Encounters with black holes in globular
clusters are usually dominated by gravitational focusing instead of
the pure geometrical cross section; this is true within $\sim 100$AU
of a 10 $M_\odot$ black hole, where we have assumed a velocity
dispersion of 10~km~s$^{-1}$ for the interlopers (see Miller
\& Hamilton 2002). In this limit the encounter time is
\begin{equation}
\tau_{\rm enc}\approx 6\times 10^5
n_6^{-1}(1~{\rm AU}/a_2)(10\,M_\odot/M_2)~{\rm yr}\; ,
\end{equation}
where the number density of stars in the core of the globular is
$10^6n_6$~pc$^{-3}$. Note that it is the semimajor axis of the
outermost object, $m_2$, that sets the encounter time scale, because
in a stable hierarchical triple $a_2$ must be a factor of several
greater than $a_1$.

The timescale for merger by gravitational radiation for a high
eccentricity orbit is (Peter 1964)
\begin{equation}
\tau_{\rm GR}\approx 5\times 10^{17}\left(M_\odot^3\over{M_1^2\mu_1}
\right)\left(a_1\over 1~{\rm AU}\right)^4\epsilon^{7/2}\,{\rm yr}\; .
\end{equation}
The steep dependence on eccentricity means that shrinkage of the orbit
is dominated by the time spent near maximum eccentricity.  Assuming
that $\tau_{\rm GR} \gg \tau_{\rm evol}$ so that orbital decay occurs
over many Kozai oscillation cycles, one finds that the fraction of
time spent near $e_{\rm max}$ is of order $\epsilon_{\rm min}^{1/2}$
(Innanen et al. 1997, equation~[5]), so that $\tau_{\rm GR}\approx 5\times
10^{17}\left(M_\odot^3\over{M_1^2\mu_1} \right)\left(a_1\over 1~{\rm
AU}\right)^4\epsilon_{\rm min}^3\,{\rm yr}$.  The condition for merger
before an encounter is then simply $\tau_{\rm GR}<\tau_{\rm enc}$.

\vspace {-3.3cm}
\centerline{\psfig{file=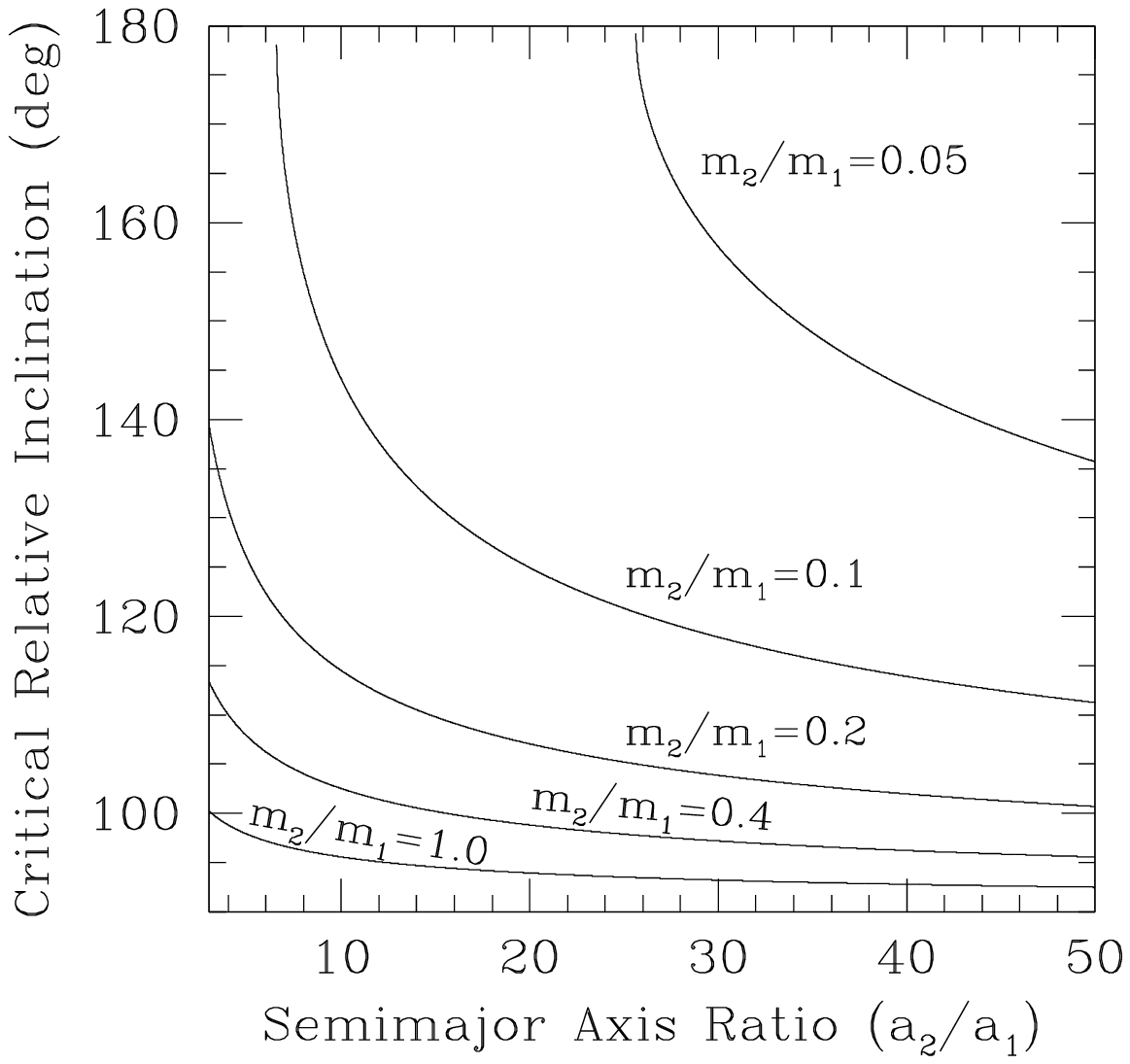,height=15.0truecm}}
\vspace {-3.8cm}
\figcaption[]{\footnotesize Critical relative inclinations for
evolution from $e\approx 0$ to $e=1$ in the Newtonian case; attaining $e=1$
requires $\alpha=\beta$ in equation~(\ref{eqn:cosI}).  Here the inner binary
is composed of equal mass stars, $m_0=m_1$, and the labels on the
curves indicate the mass ratio $m_2/m_1$.  As the fraction of the
total angular momentum supplied by the tertiary increases (larger
$\beta$ and therefore larger $m_2/m_1$ or $a_2/a_1$), the critical
inclination trends toward 90$^\circ$.}
\medskip

Note that in the Newtonian case $\theta_{\rm PN}\equiv 0$, all systems
with the same masses, $b_2/a_1$, and $I_0$ are dynamically identical, in that
the maximum eccentricity does not depend on the individual values of
$b_2$ and $a_1$.  The introduction of post-Newtonian precession breaks
this scaling.  If $b_2/a_1$ is fixed, then $\theta_{\rm PN}\propto
a_1^{-1}$ and therefore the maximum eccentricity attained is given by
$\epsilon_{\rm min}\propto a_1^{-2}$ (cf. equation~[\ref{eqn:emin}] for the
restricted problem).  The merger time is $\tau_{\rm GR}\propto
a_1^4\epsilon_{\rm min}^3\propto a_1^{-2}$. That is, a {\it wider}
binary can be pushed to higher eccentricities, and actually merge
faster, than a closer binary.  Note, however, that the solid angle for
this orientation is proportional to $\theta_{\rm PN} \propto
a_1^{-1}$, because the optimum angle is usually close to $\pi/2$, so
the solid angle is proportional to the cosine of the
inclination. Therefore, if binary-binary interactions leave the binary
and tertiary inclinations randomly oriented with respect to each other
then a smaller fraction of wide binaries will fall into the optimal
orientation.  Qualitatively this means that as the binary is hardened
by various interactions, every time a triple is formed it has a chance
to push the eccentricity high enough that the binary merges before the
next encounter.  The smaller the system, the larger the probability of
such an orientation, because both the solid angle and the encounter
time are larger.

One way to quantify the probability of merger through the increase of
eccentricity is to plot, as a function of the semimajor axis of the
inner binary, the range of relative inclinations such that merger
occurs before the next encounter of a field black hole with the
tertiary (which, being on a wide orbit, will interact before the inner
binary will on average).  In Figure~2, we assume three $10\,M_\odot$
black holes, with a given $a_1$ and $a_2$.  From $a_2$ and an assumed
number density of stars in the cluster ($n=10^6$~pc$^{-3}$), we
compute the average time $\tau_{\rm enc}$ to the next encounter within
a distance $a_2$ of the system. We then determine the range of
initial inclinations $I$ such that $\tau_{\rm GR}<\tau_{\rm enc}$, by
solving for $\epsilon_{\rm min}$ using equation~(\ref{eqn:Wgr}) with the
initial conditions $e_1 = e_2 = 0.01$ and $\omega=0$.  Note that for
wider tertiary orbits, the total angular momentum of the system is
dominated more by the tertiary (larger $\beta$), and hence the
relative inclination that gives the smallest possible $\epsilon_{\rm
min}$ is closer to $90^\circ$ (see Figure~1).  If a single Kozai
oscillation cycle is longer than $\tau_{\rm enc}$ the system never
attains the required high eccentricity.  This causes the cutoff in the
$a_2=10a_1$ and $a_2=20a_1$ curves in Figure~2; similar cutoffs exist
at $a_1>10$~AU for the remaining two curves.

\vspace {-3.3cm}
\vbox{\centerline{\psfig{file=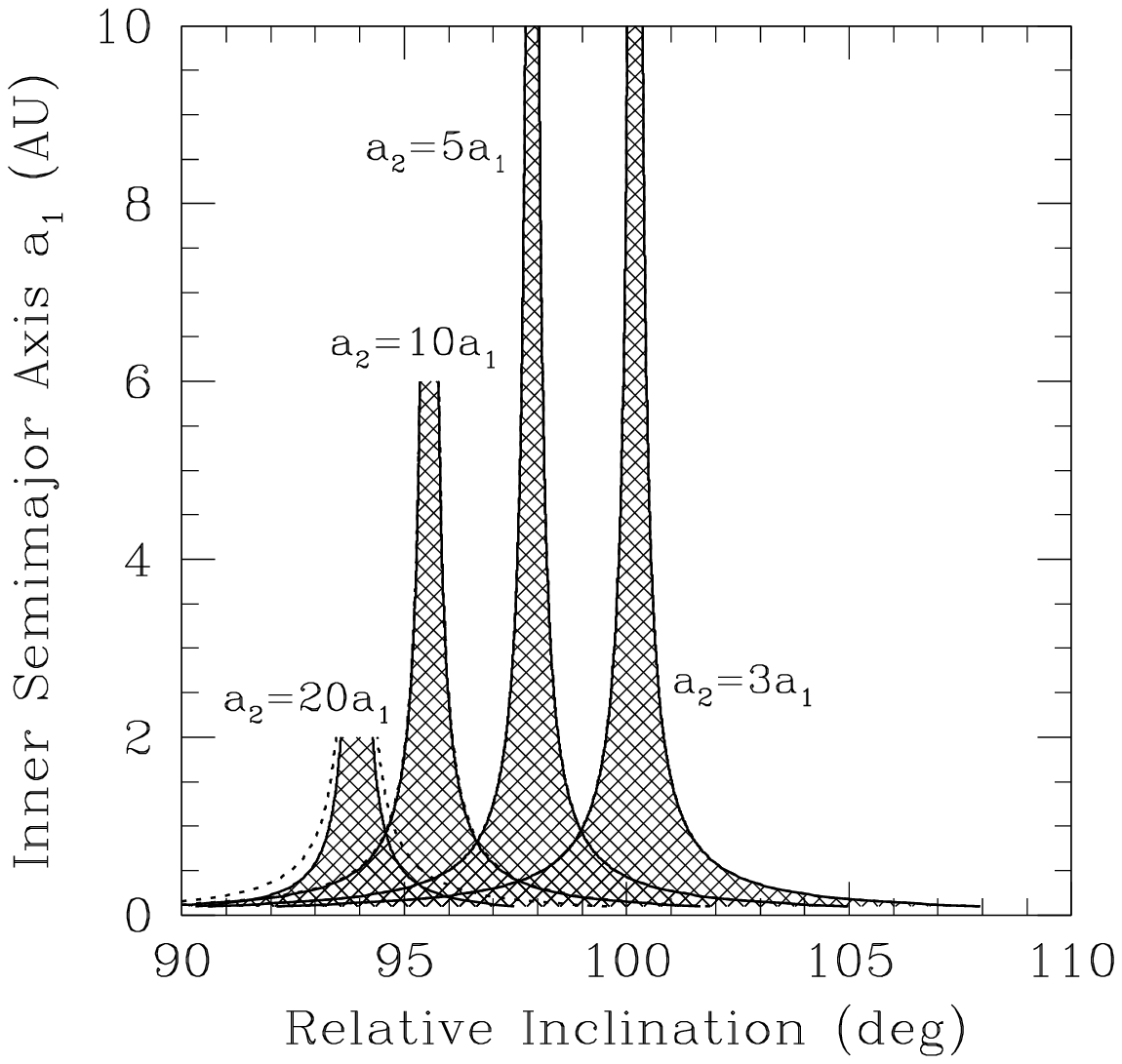,height=15.0truecm}}
\vspace {-3.8cm} \figcaption[]{\footnotesize Inclination ranges for
merger by gravitational radiation. For this graph, we assume that all
three black holes have mass $M=10\,M_\odot$, and we assume a globular
core number density $n=10^6$~pc$^{-3}$ for calculating $\tau_{\rm
enc}$. The shaded regions indicate ranges of the relative inclination
for which mergers will occur for each of four value of the semimajor
axis ratio $a_2/a_1$: 3, 5, 10, and 20. The peaks, which occur at the
locations predicted by the bottom curve in Fig.~1, are truncated where
the time to increase the eccentricity of the inner binary is greater
than the mean time to an encounter ($\tau_{\rm evol} < \tau_{\rm
enc}$). For comparison, the dotted lines are the boundaries
of the regions if general relativistic precession is suppressed; only
for $a_2=20a_1$ is there a noticeable difference.}}

\section{Conclusions}

The level of importance of the Kozai mechanism depends on several
factors including: i) details of the interactions between two
binaries, ii) details of the interactions between a triple, and either
a binary or a single star, and iii) the fraction of black holes in
binaries, which in turn relies on the iv) dynamics of the cluster
itself.  Understanding these interactions statistically will require
extensive long-term simulations.  However, the Kozai mechanism has the
potential to be the dominant process in the interactions of
stellar-mass black holes in globulars, if most such black holes are
in binaries.  When only three-body interactions are considered, very
few black holes are retained by the clusters (only 8\% in the
simulations of Portegies Zwart \& McMillan 2000).
This occurs because the same processes that harden a binary toward
an eventual merger also impart velocity kicks on the binary that
ultimately eject it from the globular before it can merge.
In contrast, the majority of black holes can be retained if
binary-binary interactions dominate.  

For example, suppose that a third of those interactions produce stable
triples.  Subsequent interactions of the tertiary with field stars will
change its eccentricity and semimajor axis. If the pericenter distance of
the tertiary is less than a few times $a_1$, then the triple system
becomes unstable, normally by ejecting its least massive member.
Suppose that there are typically $\sim$2 encounters before the
triple is disrupted in this way, and that each encounter of the tertiary
that does not create an unstable triple produces a new relative inclination
$I$ that is drawn from a uniform distribution in $\cos I$.  Suppose also
that every time the inner binary interacts strongly its semimajor axis is
decreased by  $\sim 20$\%
(typical for strong interactions of three equal-mass objects; see, e.g.,
Heggie 1975; Sigurdsson \& Phinney 1993).  Then, in an $n=10^6$~pc~$^{-3}$
cluster there is a $\approx 50$\% chance that the inner binary will merge
before it hardens to $a_1\approx 0.2$~AU, at which point the binary recoil
velocity $v_{\rm recoil}$ exceeds the $\sim$50~km~s$^{-1}$ escape speed
typical of the cores of globulars (Webbink 1985).  In an
$n=10^5$~pc~$^{-3}$ cluster, encounters are less frequent and the fraction
rises to $\approx 70$\%.

Thus, depending on the binary fraction and other properties of black
holes in globulars, the majority of black holes could merge before
being ejected, and growth of intermediate-mass black holes in globulars
may proceed naturally even if no black hole is formed with $M>10\,M_\odot$.
This could influence stellar dynamics in the core, and the
gravitational wave signals from globulars, and should 
be included in future simulations.

\vskip -0.4truecm
\acknowledgements

This work was supported in part by NASA grant NAG 5-9756 and by
NSF grant 5-23467.

\end{document}